\documentclass[prl,reprint,aps,showpacs,merge]{revtex4-1}
\usepackage{times,amsmath,graphicx,latexsym}
\begin{document}
\title{Non-universal Casimir Effect in Saturated Superfluid $^4$He
Films at T$_\lambda$}

\author{John B. S. Abraham}
\author{Gary A. Williams}
\affiliation{Department of Physics and Astronomy, University of California, Los Angeles, CA 90095}


\author{Konstantin Penanen}
\affiliation{Jet Propulsion Laboratory, California Institute of Technology, Pasadena, CA 91109}

\date{\today}

\begin{abstract}
Measurements of Casimir effects in $^4$He films in the vicinity of the bulk superfluid transition temperature $T_\lambda$ have been carried out, where changes in the film thickness and the superfluid density are both monitored as a function of temperature.  The Kosterlitz-Thouless superfluid onset temperature in the film is found to occur just as the Casimir dip in the film thickness from critical fluctuations becomes evident. Additionally, a new film-thickening effect is observed precisely at $T_\lambda$ when the temperature is swept extremely slowly. We propose that this is a non-universal Casimir effect arising from the viscous suppression of second sound modes in the film.\end{abstract}


\maketitle

\noindent Copyright 2013.  All rights reserved.

Saturated liquid $^4$He films, in contact with the bulk liquid, form an illuminating model system as a condensed matter analogue of the electromagnetic Casimir effect.   Thermal fluctuations in the film are limited by the finite thickness of the film, leading to a free-energy difference with the unlimited fluctuations in the bulk \cite{kretch}.  This causes a change in the film thickness as the atoms move to minimize the free energy.  The equilibrium thickness is determined by an energy-balance relation per atom of mass $m$ for a film of thickness $d$ at a height $h$ above the bulk helium surface,
\begin{equation}
mgh = U_{vdW}  + K_{crit}  + K_G  + K_2 \quad , 
\end{equation}
where the substrate van der Waals interaction $U_{vdW}$ at the film surface \cite{cole1988}, including retardation effects, is given by $(\gamma_0 /d^3 )(1 + d/d_{1/2} )^{ - 1} $.
This is the main term determining the film thickness, with $\gamma_0$ the van der Waals constant equal to 3.59$\times10^{-13}$ 
erg\,\AA$^3$ for a Cu substrate, and $d_{1/2}$ = 193 \AA\ .  

The fluctuation-induced Casimir forces $K$ then produce small additional shifts $\Delta d$ as the temperature is swept near the bulk superfluid transition temperature $T_{\lambda}$.  The force $K_{crit}$ comes from critical fluctuations near the transition temperature, i.e.\,the vortex loops in the model of Ref.\,\cite{gwcasimir}.  Finite-size scaling theories \cite{fisher1978,danchev1996,krech1999} show that $K_{crit} = V k_B T_{\lambda} \theta (x)/ d^3$ where $\theta (x)$ is a universal scaling function of the variable $x=t(d/\xi_0)^{1/\nu}$ with $t = (T - T_{\lambda}) / T_{\lambda}$, $\nu$ = 0.6717, $\xi_0$ =  1.43 \AA\ the amplitude of the coherence length above $T_{\lambda}$ (using the normalization of Ref.\,\cite{dietrich2009}), and $V$ = 45.8 \AA $^3$ the volume per helium atom.  Experiments \cite{hallock1989,chan1999,chan2004,chan2006} have observed a dip in the film thickness corresponding to these critical fluctuations in agreement with simulations \cite{hucht2007}, and data collapse of  films of different thickness confirmed the form of the scaling.  One factor not well determined in experiments to date is the location of the Kosterlitz-Thouless (KT) superfluid transition relative to the dip in film thickness.  A quartz microbalance oscillator technique at 5 MHz \cite{chan2004} seemed to show the transition as occurring at a temperature somewhat below the bottom of the dip, but the exact location could not be pinpointed because the high frequency greatly broadens the KT transition \cite{reppy,*hieda,*ahns}. 

The term $K_{G}$ in Eq.\,1 refers to the free energy difference between bulk and film due to thermally excited Goldstone modes\cite{kardar1992}.  It was originally proposed that such modes should include the second sound mode in the case of liquid 
helium \cite{chan2006,kardar1999,rudnick2004}, but since we question this we include instead a separate term $K_{2}$ for second sound, to be discussed below.  For Goldstone modes such as spin waves in the XY model the free energy difference arises because standing waves form in the direction perpendicular to the film surface, eliminating modes that can still propagate in the bulk.  This gives $K_{G} = V k_{B}T \zeta (3)/(8 \pi d^3)$, a result well confirmed in computer simulations \cite{hucht2007,dietrich2009,hasenbusch2009,hasenbusch2010} and analytic calculations \cite{dohm2013} for XY spin waves.  The experiments in helium did find a difference in film thickness between low temperatures and above $T_{\lambda}$ similar to but larger than this prediction \cite{chan1999,chan2006}; however a later theory \cite{rudnick2004} showed that this was primarily due to energy differences in the waves at the free surface of the film and bulk, resulting in a thinning magnitude nearly double the the expression for $K_{G}$ above.

We doubt very much that second sound propagates at all in saturated helium films.  Second sound involves counterflow of the normal fluid of viscosity $\eta$ and density $\rho_n$, and in a film this will halt any flow at angular frequencies $\omega$ where the viscous penetration depth $l_v = (2 \eta/ \rho_n \omega)^{1/2}$ is larger than the film thickness \cite{bergman1971}.  For a film with $d$ = 300 \AA\ near $T_{\lambda}$ this is all frequencies below $\omega/2\pi$ = 6 MHz.  Any possible modes at higher frequencies in the film would necessarily be waveguide modes reflecting from the free surface, and because second sound involves temperature oscillations these would excite sound waves in the helium vapor, leading to very strong attenuation from the dissipative Onsager reciprocity relations for the evaporation-condensation process \cite{bergman1971,williamstwophase}.  Such waveguide modes coupled to vapor in bulk helium have been observed to be strongly attenuated (Q values of order 5) at only a few hundred Hz \cite{williamstwophase}, so we believe there is almost no probability they could propagate at MHz frequencies in the film.

If there is no second sound in the film, then $K_{2}$ in Eq.\,1 will just be equal to $F_2 V$, where $F_2$ is the free energy per unit volume of second sound in the bulk superfluid.  Second sound is well known to propagate in the bulk to within microkelvins of $T_{\lambda}$ \cite{ahlers1973,greytak1977}, but it then ceases propagation above the superfluid transition, which will cause $K_2$ to drop to zero.  This means from 
Eq.\,1 that there must be a step increase at $T_{\lambda}$ in the film thickness, which to linear order in $\Delta d /d$ is 
\begin{equation}
\frac{{\Delta d}}
{d} = \frac{{ - F_2 (T_\lambda  ) V}}
{{3\;U_{vdW} }}\frac{{\left( {1 + \frac{d}
{{d_{{\raise0.5ex\hbox{$\scriptstyle 1$}
\kern-0.1em/\kern-0.15em
\lower0.25ex\hbox{$\scriptstyle 2$}}} }}} \right)}}
{{\left( {1 + \frac{4}
{3}\frac{d}
{{d_{{\raise0.5ex\hbox{$\scriptstyle 1$}
\kern-0.1em/\kern-0.15em
\lower0.25ex\hbox{$\scriptstyle 2$}}} }}} \right)}}\quad .
\end{equation}

We report here the experimental observation of just such a step increase in the thickness at precisely $T_{\lambda}$, but which only becomes apparent for an extremely slow sweep near the transition.  Also, in studies of the critical-fluctuation Casimir effect we have used third sound measurements on the same films to determine the KT onset temperatures, and find that the onset occurs at the very beginning of the dip in thickness.

\begin{figure}[t]
\begin{center}
\includegraphics[width=0.45\textwidth]{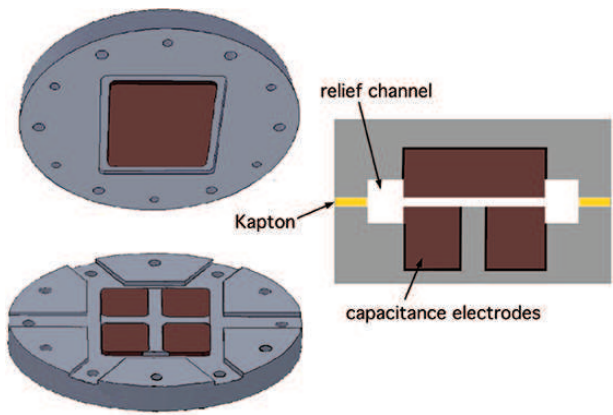} 
\end{center}
\caption{Schematic of the capacitor electrodes. (Color online)}
\label{fig1}
\end{figure}

 Our experimental cell is mounted in a vacuum space inside a conventional 1 K pot cryostat.  There are three intermediate thermal stages between the 1 K pot held at 1.9 K and the experimental stage, which minimizes thermal gradients.  The temperature of the experimental stage is regulated by applying feedback heat to the intermediate stages using PID temperature control, while the cell itself is not directly regulated.  The cell temperature is measured with a germanium thermometer attached to the outside of the cell, with a resolution of about 10 $\mu$K.  The position of $T_{\lambda}$ can be determined to within 30 $\mu$K by the change in the temperature drift of a carbon thermometer at the end of a stainless tube extending from the bottom of the cell.
All thermal connections are made with gold plated copper press joints to maximize the thermal conductivity. The cell fill line capillary is connected to the bottom of the cell through a pressure actuated cold valve, so that the cell can be isolated from other parts of the cryostat during measurement. 
\begin{figure}[b]
\begin{center}
\includegraphics[width=0.48\textwidth]{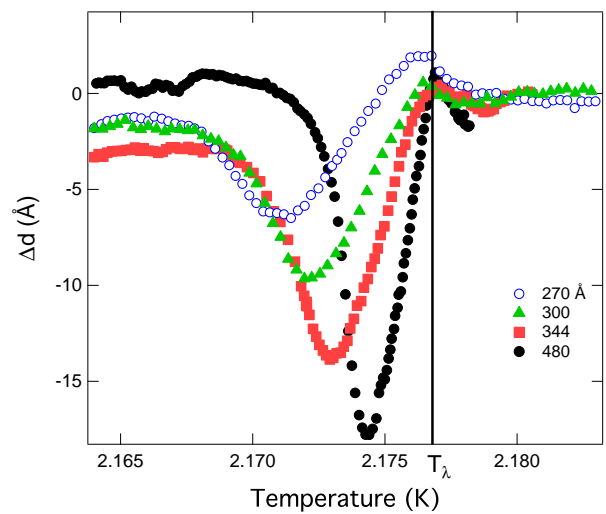} 
\end{center}
\caption{ Change in the film thickness versus temperature. (Color online)}
\label{fig2}
\end{figure}

Our capacitor substrates (shown schematically in Fig.\,1) are fashioned from OFHC copper, which after assembly in copper holding plates with Stycast 1266 epoxy, are diamond-turned to give a surface roughness of less than 10 nm, as checked with an atomic force microscope.  The capacitor plates are cleaned and assembled in a microfabrication facility and clean room to minimize any possible dust.  The lower capacitor plate is partitioned into four rectangular quadrants of size 0.9 $\times$ 1.2 cm, in order to probe the He films with third sound.  2 mm grounding strips minimize any cross talk between the quadrants.  A relief channel 1 mm deep and 2.5 mm wide is machined out of the region surrounding both the upper and lower electrodes to prevent any capillary filling. The gap is set by placing a spacer of 60 $\mu$m thick Kapton film between the substrate holders, touching only at the outer edge of the relief channel.  Six channels in the lower holder allow the helium film and vapor to enter freely.  

The empty cell capacitance of a quadrant is approximately 15 pF.  An identical electrode assembly is bolted to the outside of the sealed cell in the vacuum, and forms a reference capacitor for a ratio transformer bridge.  The ratio transformer at room temperature is thermally insulated and temperature-regulated to 0.1 C. The thickness of the film is measured from the capacitance ratio by subtracting out the vapor contribution to the cell capacitance, using the same methods as Refs.\,\cite{chan1999,chan2006}.  Third sound is generated in the film by applying an oscillating electric field to one of the quadrants, sweeping slowly from 3-20 Hz, and detecting resonant modes by the capacitance change of the diagonal quadrant \cite{packard}.  The equilibrium film thickness well below the $\lambda$ point is actually most accurately measured with the third sound speed \cite{abrahamthesis}.
\begin{figure}[t]
\begin{center}
\includegraphics[width=0.5\textwidth]{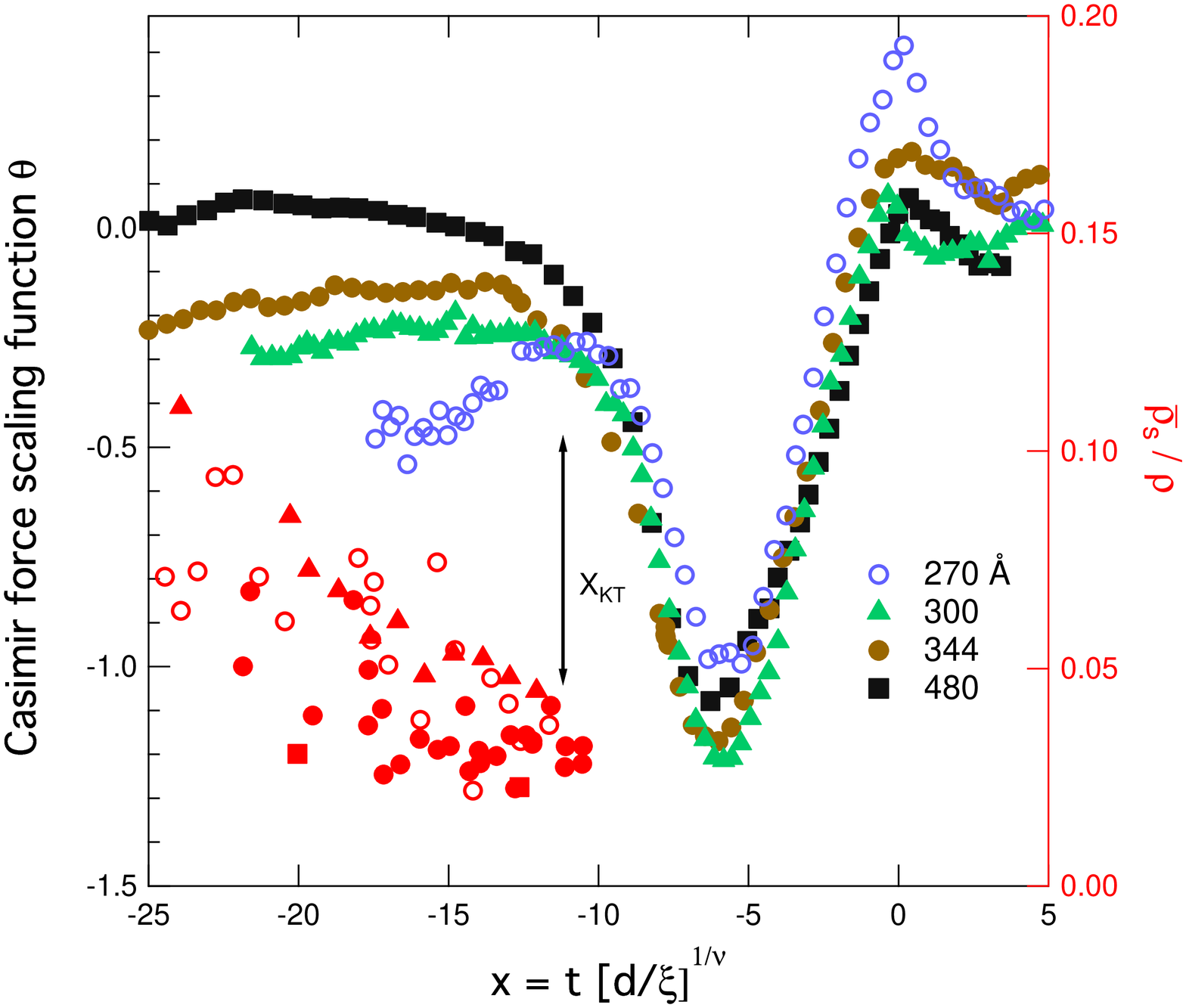} 
\end{center}
\caption{Casimir force scaling function from the data of Fig.\,1 (left axis), and average superfluid density (right axis). (Color online)}
\label{fig3}
\end{figure}

Figure 2 shows the change in film thickness for several films near $T_{\lambda}$ at a temperature sweep rate about 400 $\mu$K/hr. The Casimir force $K_{crit}$ from critical fluctuations is readily apparent, causing a dip in the film thickness when the coherence length equals the film thickness, in good agreement with the previous measurements.  Figure 3 shows the Casimir force scaling function, with the same good data collapse as also seen previously \cite{chan1999,chan2006}. At the dip minimum we find $x_{min} = -6.0(5)$ and $\theta_{min} = -1.2(1)$, compared with $x_{min} = -5.7(5)$ and 
$\theta_{min} = -1.30(3)$ found previously.  This can also be compared with Monte Carlo simulations on XY spins with free boundary conditions at the top and bottom boundaries \cite{dietrich2009,hasenbusch2009,hasenbusch2010}, which find $x_{min} = -5.3(1)$ or -5.43(2), and $\theta_{min} = -1.35(3)$ or -1.396(6).

Our ability to monitor third sound in these films allows us to determine the Kosterlitz-Thouless onset relative to the dip in thickness.  The superfluid density averaged across the film can be extracted from the third sound speed, shown as the red symbols in Fig.\,3 plotted on the right-hand axis.  The signal can no longer be detected past the last points plotted, and at that point the value of $\bar \rho _s d/T_{KT}$ averaged over the different films is 4.7$\pm$1.5 ng/cm$^2$ K, only somewhat above the universal KT value of 3.5 ng/cm$^2$ K.  This occurs at a value $x_{KT}$ = -11.1(4) shown by the arrow. These very low frequency (4-5 Hz) onset values  of $T_{KT}$ are consistent with previous results in films near the lambda point \cite{gaspariniRMP}.  The onset occurs at the very beginning of the dip in thickness, a result predicted in the vortex-loop theory \cite{gwcasimir} (though only for periodic boundary conditions).  The Monte Carlo simulations \cite{dietrich2009} give the KT transition at  $x_{KT}$ = -7.6, more in the middle of the dip, but this is due to the fact that XY spins have a higher vortex core energy than found for helium films \cite{cho1995}, and the higher core energy  shifts $T_{KT}$ to higher temperatures.  

We were initially puzzled by the relative maximum in the film thickness near the transition temperature $T_{\lambda}$ seen in Figs.\,2 and 3, which was also observed in the quartz microbalance experiment \cite{chan2004}.  We found, however, that this feature was quite dependent on the temperature sweep rate.  When we increased the sweep rate to about 2 mK/hr,  we found only a smooth increase through $T_{\lambda}$ to a constant value, very similar to the behavior seen in Ref.\,\cite{chan1999}.  We decided to sweep as slowly as we could near 
the transition, starting a few hundred  $\mu$K below $T_{\lambda}$ and taking upward steps of 20 $\mu$K, equilibrating for 3-5 hours at each point for the temperature and capacitance to stabilize.  The results are shown in Figure 4 for three film thicknesses.   Below $T_{\lambda}$ there is little variation over this very restricted range, but right at $T_{\lambda}$ a sudden step increase in the thickness occurs, with the magnitude of the increase a strong function of the equilibrium film thickness.  We suspended the temperature stepping once the thickness started to increase; for example with the 480 \AA\ film the increase was followed for about 15 hours before it started to level off and the temperature steps were resumed.  By checking different quadrants we found the same value of the step increase, showing that the film is uniform across the cell and that this effect is not due to capillary condensation, which would tend to localize the thickness change to one quadrant.  We also checked that this effect is present for temperature sweeps both up and down in temperature through the lambda point, shown in Fig.\,5 for two sweeps of the 344 \AA\ film, taken a week apart.  To within our temperature and thickness resolution, these sweeps show that the step behavior is quite reproducible.  
\begin{figure}[b]
\begin{center}
\includegraphics[width=0.5\textwidth]{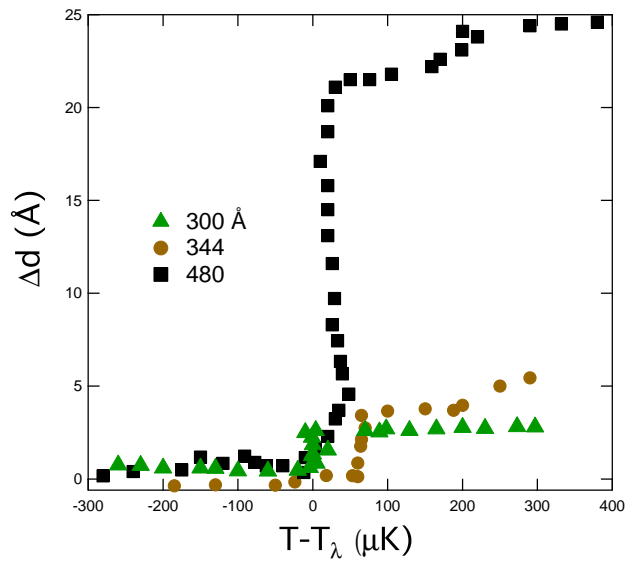} 
\end{center}
\caption{Change in film thickness observed at the superfluid transition for extremely slow temperature sweeps.  (Color online)}
\label{fig4}
\end{figure}

Figure 6 shows the relative change in film thickness $\Delta d/d$ at the transition as a function of $d$, which increases rapidly.  This is quite consistent with the rapid increase predicted by Eq.\,2, since $U_{vdW}$ is a strong function of $d$.  The solid line in Fig.\,6 is a fit to a rootfinder solution of the full Eq.\,1,  with the second-sound free energy $F_2$ as the only free parameter, yielding  $F_2$ = -3.5 erg/cm$^3$.  To our knowledge, there has never been a calculation or measurement of the free energy of second sound in bulk helium to compare with our value for $F_2$ at $T_{\lambda}$.  This would require a detailed theory of the properties of thermally excited second sound near $T_{\lambda}$ at MHz frequencies, a regime where the wavelength is smaller than the correlation length, well outside the low-frequency hydrodynamic regime.  There has been some theoretical work in this area \cite{halperin1976,siggia1976}, but comparison to experiment has only been at the single wavelength investigated in the light-scattering experiments \cite{greytak1977}.
\begin{figure}[t]
\begin{center}
\includegraphics[width=0.5\textwidth]{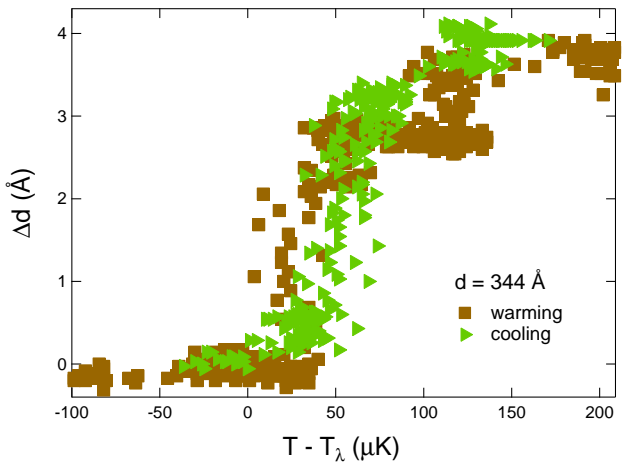} 
\end{center}
\caption{Change in the 344 \AA\ film thickness for sweeps warming and cooling through the transition.  (Color online)}
\label{fig5 }
\end{figure}

An initial calculation of the free energy can be attempted using the Debye approximation commonly used for phonons in solids.  We assume that the angular Debye cutoff frequency $\omega_D$ is the highest frequency where the mode propagates, e.\,g.\, where the quality factor $Q \approx$ 1.  In the limit 
$\hbar \omega/k_B T \ll 1$ (a good approximation since $c_2$ is relatively small), the expression for the free energy is 
\begin{eqnarray}
 F_2  = \frac{{k_B T}}{{2\pi ^2 c_2^3 }}\int_0^{\omega _D } {\omega ^2 \ln \left( {\frac{{\hbar \omega }}{{k_B T}}} \right)} \;d\omega  \\ \nonumber
 \quad \, =  - \frac{{k_B T}}{{2\pi ^2 }}\left( {\frac{{\omega _D }}{{c_2 }}} \right)^3 \left[ {\frac{1}{9} - \frac{1}{3}\ln \left( {\frac{{\hbar \omega _D }}{{k_B T}}} \right)} \right]. \\ \nonumber
\end{eqnarray}
In evaluating this expression for T very close to $T_{\lambda}$, the value of $c_2$ at MHz frequencies does not go to zero, due to finite-wavenumber and finite-frequency broadening (seen in the light-scattering measurements \cite{greytak1977} and known from theory \cite{halperin1976,siggia1976,gwk,*gwfreq}), but becomes roughly constant starting about 50
$\mu K$ from $T_{\lambda}$.  The light-scattering measurements very close to $T_{\lambda}$ are only available at high pressure (23.1 and 28.5 bars), but a rough extrapolation to the saturated vapor pressure of the films using low-frequency  measurements \cite{ahlers1973} gives an estimate for $c_2$ at $T_{\lambda}$ of 80 cm/s.  Inserting this value and the fitting result of the step data, $F_2(T_{\lambda}) \approx$ -3.5 erg/cm$^3$ into Eq.\,3 we find a value of $\omega_D$/2$\pi \approx$ 6 MHz.  This is a fairly reasonable order of magnitude estimate, since the Brillouin line in the light scattering data near $T_{\lambda}$ at 1.93 bars \cite{greytak1974} already shows strong attenuation, with the Q $\approx$ 3 at a center frequency of 4 MHz at 500 $\mu K$ from the transition.  However, the frequency dependence of the attenuation in this non-hydrodynamic regime is not well known.  There is certainly a need for  a more detailed theory of the free energy of second sound close to $T_{\lambda}$. 
\begin{figure}[t]
\begin{center}
\includegraphics[width=0.45\textwidth]{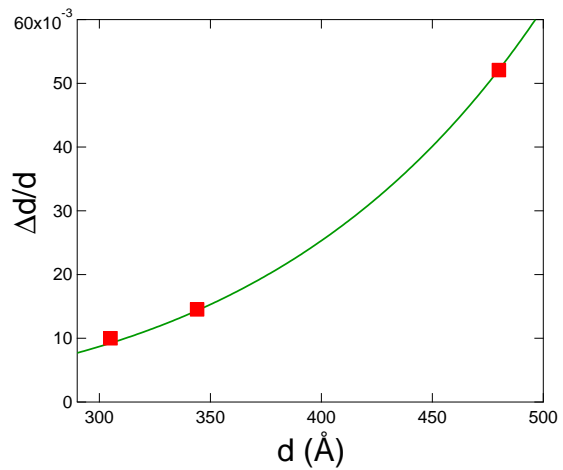} 
\end{center}
\caption{Relative jump in the film thickness at $T_{\lambda}$.  The solid line is a fit to Eq.\,1, giving $F_2(T_{\lambda})$ = -3.5 erg/cm$^3$. (Color online)}
\label{fig6}
\end{figure}

The long relaxation time for the appearance of the sharp step at $T_{\lambda}$ probably arises from the complete lack of superfluidity in both the film and the bulk.  The only means of mass transport to change the film thickness is then through the vapor, and this is a very slow mass diffusion process, particularly since the vapor in the 60 $\mu$m channel is also viscously constrained.  It is very easy to miss the step (as we did numerous times early in this data run); the key procedural difference in observing the effect was to halt the (already very slow) temperature sweep at the first sign of even a slight change in the capacitance \cite{comment}.  We did not observe any noticeable change in the mass relaxation rate in the film at and above $T_{KT}$ for different sweep speeds (similar to Refs.\,\cite{chan1999,chan2006}), probably because there is still finite-frequency superfluidity in the film \cite{reppy,*hieda,*ahns}, which finally disappears only at $T_{\lambda}$.

We note that the first sound mode very likely does not play any role in this Casimir effect in helium films.  First sound is known to propagate to very high frequencies in both the bulk \cite{neutron} and in films \cite{anderson1972}.  For such wavelengths smaller than the film thickness there will be little difference between the free energy of the film and the bulk, and no change at the lambda point since the modes exist both below and above $T_{\lambda}$.

In summary, we identify the onset of the KT transition in saturated helium films to occur right at the start of the Casimir dip in film thickness.  In addition we observe a new step increase in the film thickness at $T_{\lambda}$ that we attribute to the abrupt cessation of second sound in the bulk liquid.  Since this Casimir effect is related to the unique properties of second sound in helium (the viscous suppression of the mode in films) and involves material-dependent quantities such as $c_2$, it will not be a universal property of XY superfluids.
\begin{acknowledgments}
We thank Joseph Rudnick for very useful discussions, Hong-Wen Jiang for the use of the AFM, and Vincent Wong for help with data analysis.  A portion of this work was carried out at the Jet Propulsion Laboratory, California Institute of Technology, under a contract with the National Aeronautics and Space Administration, and was also supported in part by the National Science Foundation, grant DMR-0906467.
\end{acknowledgments}
\providecommand{\noopsort}[1]{}\providecommand{\singleletter}[1]{#1}%

\end{document}